# The interpretation of endobronchial ultrasound image using 3D convolutional neural network for differentiating malignant and benign mediastinal lesions

Ching-Kai Lin[1,2,3,‡], Shao-Hua Wu[4,‡], Jerry Chang[4], Yun-Chien Cheng[4,*],

[1] Department of Medicine, National Taiwan University Cancer Center, Taipei, Taiwan

[2] Department of Internal Medicine, National Taiwan University Hospital, Taipei, Taiwan

[3] Department of Internal Medicine, National Taiwan University Hsin-Chu Hospital, Hsin-Chu, Taiwan

[4] Department of Mechanical Engineering, College of Engineering, National Yang Ming Chiao Tung University, Hsin-Chu, Taiwan

[‡]The authors contributed equally to this work.
[*]Email: yccheng@nycu.edu.tw

**Abstract**

The purpose of this study is to differentiate malignant and benign mediastinal lesions by using the three-dimensional convolutional neural network through the endobronchial ultrasound (EBUS) image. Compared with previous study, our proposed model is robust to noise and able to fuse various imaging features and spatiotemporal features of EBUS videos. Endobronchial ultrasound-guided transbronchial needle aspiration (EBUS-TBNA) is a diagnostic tool for intrathoracic lymph nodes. Physician can observe the characteristics of the lesion using grayscale mode, doppler mode, and elastography during the procedure. To process the EBUS data in the form of a video and appropriately integrate the features of multiple imaging modes, we used a time-series three-dimensional convolutional neural network (3D CNN) to learn the spatiotemporal features and design a variety of architectures to fuse each imaging mode. Our model (Res3D_UDE) took grayscale mode, Doppler mode, and elastography as training data and achieved an accuracy of 82.00% and area under the curve (AUC) of 0.83 on the validation set. Compared with previous study, we directly used videos recorded during procedure as training and validation data, without additional manual selection, which might be easier for clinical application. In addition, model designed with 3D CNN can also effectively learn spatiotemporal features and improve accuracy. In the future, our model may be used to guide physicians to quickly and correctly find the target lesions for slice sampling during the inspection process, reduce the number of slices of benign lesions, and shorten the inspection time.

## 1. Introduction

Lung cancer is the most common cause of cancer-related death worldwide [1]. Accurate diagnosis and staging are essential to devising a proper treatment plan [2]. Endobronchial ultrasound-guided transbronchial needle aspiration (EBUS-TBNA) is a relatively new and minimally invasive procedure for approaching mediastinal and hilar lesions. Previous publications have confirmed its high accuracy in diagnosis and staging of lung malignancy [3-6]. However, several possible targets are frequently found in the same station during the EBUS-TBNA. Distinguishing the most likely target lesion is a critical issue for determining the diagnostic yield of the procedure.

Some sonographic patterns have been used to differentiating benign from malignant lesions. The grayscale mode imaging patterns (size, shape, margin, echogenicity, central hilar structure, coagulation necrosis sign, matting, calcification, and nodal conglomeration) were proved their ability to evaluate lymph node metastasis [7-9]. New sonographic techniques, such as the Doppler mode, and elastography also displayed the effectiveness for predicting the target lesions [10-12]. Though combining more imaging criteria even have better accuracy [8, 9], more procedure time are required during the procedure. Bronchoscopists even need more experience to familiarize these sonographic features. Finding an effective way to present EBUS imaging patterns during the procedures is needed.

In recent years, machine learning has become popular and convolutional neural network (CNN) [13] has made considerable progress in the field of image processing. The ResNet architecture [14] enables the CNN architecture to stack deeper and analyze more complex features. If the convolutional neural network can assist physicians in the interpretation of images during EBUS-TBNA, the target lesions can be found more quickly and accurately, and the inspection time can be shortened.

EBUS data is in the form of video. In the field of video processing, three-dimensional convolutional neural network (3D CNN) is widely used due to its ability to capture temporal information. For example, I3D [15] and Slowfast [16] have good performance on video recognition tasks.

From the aforementioned literature, we established an automatic interpretation system using 3D CNN to determine the lesions in the EBUS video. The physician can determine the benign and malignant lesions by combining our classification result and their domain knowledge. Compared to previous study, we made several improvements. First, we only segmented the time interval of lesion in the video instead of manually selecting specific pictures as training and validation data to reduce the time cost of manually selecting data in future clinical application. Second, the data used for training and validation contained noisy images, so the trained model more robust to noise. Moreover, this verification method is closer to the actual clinical situation. Third, we used 3D CNN to construct our model to obtain higher classification accuracy by retaining the temporal information in the video.

Therefore, our goal is to design a computerized automatic interpretation system that provides confidence scores in real-time to assist physicians in diagnosing benign and malignant lesions. The methods mentioned above were used to design classification models to achieve higher classification accuracy than pervious methods.

## 2. Related work

In EBUS image recognition, Jin Li et al. proposed the EBUS-net [17], based on two-dimensional convolutional neural network and can effectively classify benign and malignant lesion on EBUS images. However, the data used for training, validation, and testing were images selected by experts that fully contained the characteristics of the tumor. While in clinical application, it may cause additional burden. In addition, noisy images, probe shaking, and the different selection criteria of different physicians (especially beginners) may also affect the performance of the model.

## 3. Method

The goal of this research is to establish a deep learning model for analyzing benign and malignant lesions by ultrasound video during EBUS procedure. This study divided the data into training set and validation set by patient. We cropped the video when the lesion appeared, without additional filter. We have proposed a variety of models. The suffix of the model name is the EBUS mode used for training. U stands for grayscale mode, D stands for doppler mode, and E stands for elastography. We designed Res3D_U and Res3D_UD with 3D CNN and designed Res3D_UDE with 2D and 3D CNN. The models were all trained on the training set and evaluated on the validation set.

**3.1 Experiment process**

*3.1.1 Experiment overview*

The main purpose of this research is to design a 3D convolutional neural network to classify the benign and malignant mediastinal lesions in EBUS images.

*3.1.2 EBUS-TBNA procedure and EBUS image collection*

EBUS-TBNA was performed with the convex probe EBUS (BF-UC260FW; Olympus Co., Tokyo, Japan) and EBUS images were generated using a dedicated ultrasound processor (EU-ME2 PREMIER PLUS; Olympus Co., Tokyo, Japan). Conscious sedation with intravenous fenatyl and midazolam was administered in all patients. After inserting the convex probe EBUS via the oral route, the target lesion was detected via grayscale mode. When the target was explored, we moved the EBUS probe back and forth slowly, and then rotated clockwise and counterclockwise for scanning the whole target lesion. The Doppler mode and elastography were repeated the same protocol to gain the detail information of the target. All video data were stored via a medical video recorder system (TR2103; TWIN BEANS Co., New Taipei City, Taiwan). The dedicated 22-gauge needle (NA-201SX-4022; Olympus Co., Tokyo, Japan) were used for TBNA biopsies and all tissue samples were impregnated in 10% formalin, embedded in paraffin and stained with hematoxylin and eosin for subsequent pathological analysis. If more than one target site needs to be approached, a sequential N3-N2-N1 strategy will be followed for lymph node staging [18].

The final diagnosis of mediastinal/hilar lesion was established based on cytopathologic evidence, microbiological analyses, or clinical follow-up. Non-malignant diagnoses, which could not be determined cytopathologically or microbiologically (fibrosis/chronic inflammation), were confirmed by radiological and clinical follow-up (unchanged or decreased lesion size on the computed tomography image) at least 6 months after EBUS-TBNA.

*3.1.3 Dataset*

This study collected 67 patients and received EBUS-TBNA in the Department of Thoracic Disease, National Taiwan University Cancer Center from November 2019 to June 2020. We divided these data into training set and validation set by patient. The training set contained 45 patients, with a total of 76 benign lesions and 25 malignant lesions. The validation set contained 22 patients, with a total of 27 benign lesions and 23 malignant lesions. The study was approved by the National Taiwan University Cancer Center Institutional Review Board (IRB # 202105105RIND).

*3.1.4 Preprocessing*

The pre-processing flow is shown in Figure 1. To reduce the computational cost and remove the less important part, the EBUS images were cropped into the size of 704×576 (pixels) and recorded when the lesions appeared in the video. We adopted different pre-processing methods for the three modes: grayscale mode, doppler mode, and elastography. For grayscale mode and doppler mode, the videos of lesions were divided into several 6-second clips, with 50% overlapping between two adjacent clips. Then, each clip was sampled at a frequency of 4 times per second. At the end, 24 pictures were stacked into a three-dimensional image as a model input (slice). The imaging mode (graphic signal) of each data was recorded in a form of vector. For elastography, pictures with a larger coverage area of elastic imaging were selected as the input of the two-dimensional picture of the model through the image processing method and at most three pictures were selected for each lesion.

To our models from avoid overfitting, data augmentation was used to increase the amount of training data. The data augmentation method included 0.2 probability of horizontal flip, 0.2 probability of gaussian noise, and 0.2 probability of gaussian blur.

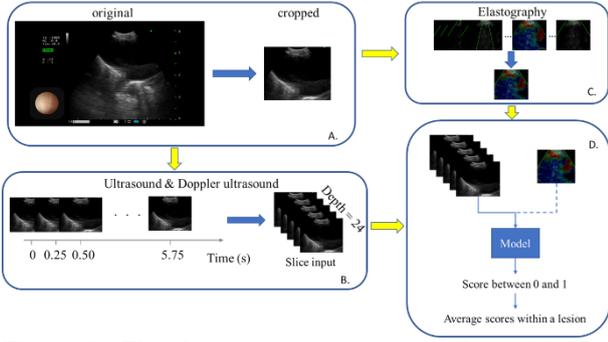

Figure 1. The data preprocessing workflow. (a) The EBUS videos were cropped to retain important information. (b) The grayscale mode and doppler mode videos in the lesion were divided into several 6-second clips, and 24 frames were sampled in each clip. (c) Selected pictures with a large area covered by elastography through image processing. (d) A multi-pathway model is designed.

### 3.2 Convolutional Neural Network

Convolutional Neural Network (CNN) is a feedforward neural network consisting of multiple convolutional layers, pooling layers, and finally a fully connected layer. Compared with other deep learning structures, the convolutional neural network has fewer parameters, effectively reduce the problem of over-fitting, and can produce better results in image and speech recognition.

### 3.3 Three-dimensional convolutional neural network

Shuiwang Ji [19] proposed a three-dimensional convolutional neural network for video action recognition. A series of pictures obtained from video sampling were stacked as three-dimensional inputs, and the features were extracted through three-dimensional convolution kernels to preserve the spatiotemporal features.

Since our data is in the form of video, each data has two spatial dimensions and a time dimension. Moreover, the variation in the ultrasound images of the lesion during EBUS procedure are important for diagnosis. Therefore, we believed that using 3D convolution as the backbone of the model can effectively improve performance.

### 3.4 ResNet

ResNet is a network architecture formed by stacking multiple convolutional layers. When the depth of network increases, the gradients vanishing and exploding problem appears, which leads to saturation of the model performance and even a decrease in accuracy. In the paper proposed by Kaiming He in 2015 [14], the concept of residual learning was used to solve these problems. Therefore, ResNet can be extended to a deep network of 50 layers or even more than 100 layers.

The goal of this study is to determine the benign and malignant lesions in EBUS images. The characteristics of lesions are complex and have high variability, so it is suitable to use the ResNet architecture as the model backbone.

### 3.5 Overall Architecture

#### 3.2.1 Res3D_UDE

The model architecture is shown in Figure 2, and the detailed parameters are shown in Table 1. To retain the temporal feature of grayscale mode and doppler mode, the model architecture was divided into a 3D path and a 2D path. The 3D path took 3D images in grayscale mode and doppler mode as input and extracted features through a 3D ResNet encoder. The 2D path took the 2D images of the elastography as input and extracted the features through the 2D ResNet encoder. The extracted 3D and 2D features were respectively passed through the fully connected layer to obtain 1000-dimensional vectors. Then the features were added element-wise. The recorded graphical signal vector was passed through the fully connected layer with the 1000-dimensional vector output as the weight and multiplied element-wise with features. Finally, the result between 0 and 1 was generated through the fully connected layer.

Each training data input consisted of a three-dimensional image (slice) in grayscale mode or doppler mode and a 2D image of elastography (if no images were available for elastography, a zero matrix was used).

To solve the gradients vanishing and exploding problem, ResNet was selected as the model backbone.

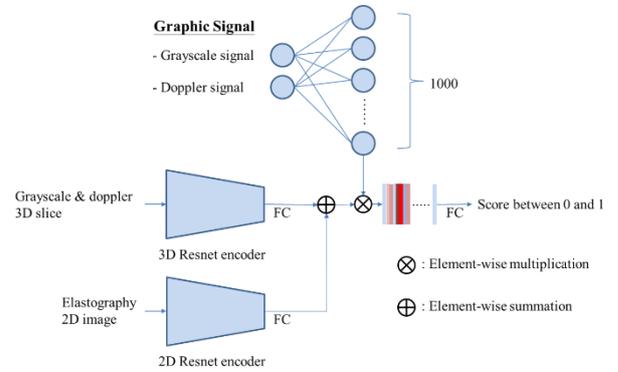

Figure 2. The Res3D_UDE network has a 3D path for extracting features from grayscale mode and doppler mode slices, and a 2D path for extracting features from elastography images. The graphic signal provided model imaging mode information.

Table 1. The overall architecture of the network. The dimensions of output were denoted by {W×H×T} for 3D pathway and {W×H} for 2D pathway. The backbone is ResNet.

| Stage | 3D pathway | 2D pathway | Output size |
|---|---|---|---|
| Raw slice | - | - | 704 × 576 × 24 |
| Conv$_1$ | Kernel $5^3$, channel 16 stride $1^3$ | Kernel $5^2$, channel 16 stride $1^2$ | 3D: 704 × 576 × 24 2D: 704 × 576 |
| Res$_1$ | Kernel $3^3$, channel 16 Kernel $3^3$, channel 16 | Kernel $3^2$, channel 16 Kernel $3^2$, channel 16 | 3D: 352 × 288 × 24 2D: 352 × 288 |
| Res$_2$ | Kernel $3^3$, channel 16 Kernel $3^3$, channel 32 | Kernel $3^2$, channel 16 Kernel $3^2$, channel 32 | 3D: 176 × 144 × 24 2D: 176 × 144 |
| Res$_3$ | Kernel $3^3$, channel 32 Kernel $3^3$, channel 64 | Kernel $3^2$, channel 32 Kernel $3^2$, channel 64 | 3D: 88 × 72 × 24 2D: 88 × 72 |
| Res$_4$ | Kernel $3^3$, channel 64 Kernel $3^3$, channel 128 | Kernel $3^2$, channel 64 Kernel $3^2$, channel 128 | 3D: 44 × 36 × 24 2D: 44 × 36 |
| Res$_5$ | Kernel $3^3$, channel 128 Kernel $3^3$, channel 256 | Kernel $3^2$, channel 128 Kernel $3^2$, channel 256 | 3D: 22 × 18 × 24 2D: 22 × 18 |
| Res$_6$ | Kernel $3^3$, channel 256 Kernel $3^3$, channel 512 | Kernel $3^2$, channel 256 Kernel $3^2$, channel 512 | 3D: 11 × 9 × 24 2D: 11 × 9 |
| Average pool, fc layer, summation, graphical signal attention | | | 1000 |
| Fc, sigmoid | | | 1 |

*3.2.2 Res3D_UD*

The model architecture is shown in Figure 3. Grayscale mode and Doppler mode were used as the model input. 3D encoder was the same as Res3D_UDE

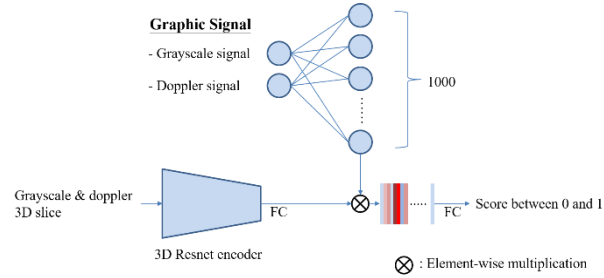

Figure 3. Res3D_UD has a 3D path extracting features from grayscale mode and doppler mode slices.

*3.2.3 Res3D_U*

The model architecture was the same as Res3D_UD, but there was no graphic signal. Only grayscale mode was used as model input.

*3.2.4 Hyperparameters*

The hyperparameters used for training in this study included the initial learning rate of 1e-4, the optimizer was stochastic gradient descent, the loss function was binary cross entropy, the parameters were updated every 12 data, the image input size was 704×576 (pixels), and the learning rate was reduced using the cosine learning rate decay.

### 3.6 Equipment

In this study, the ASUS X99-E10G WS computing platform was equipped with Intel(R) Xeon(R) CPU E5-2620 and two NVIDIA(R) GeForce(R) RTX 1080.

### 4. Result

#### 4.1 Evaluation method

There were two evaluation methods. First, we took video slice as model input. If the output result is greater than 0.5, it is regarded as malignant. The accuracy and ROC curve were analyzed. Second, we evaluated the result on lesions by averaging all the output results obtained by the slices in the same lesion. If the average score was greater than 0.5, it was considered malignant. The accuracy and ROC curve were analyzed.

The time length of grayscale mode videos were less than 6 seconds in some lesions, making it is impossible to be segmented into even a video clip. Therefore, these lesions were excluded when evaluating the Res3D_U.

#### 4.2 Comparison of different input modes

Table 2. shows our results on the validation set with different models. Figure 4. shows the ROC curve and area under curve (AUC) of these results. Res3D_UDE has 82.00% accuracy on lesion classification, while Res3D_U has only 75.76%. It is speculated that we have less grayscale mode data. When doppler mode and elastography were added to training, the accuracy can be effectively improved. In the comparison of AUC (lesion), Res3D_UDE (0.83) was slightly lower than Res3D_UD (0.86), because the model design failed to properly fuse the features of 3D and 2D. In addition, in the comparison between the result on slice and lesion, the results of the lesion were higher than the results of slice, since the model was affected by noise, which caused some slices to be misjudged. However, it did not affect the classification results of the entire lesion.

Table 2. Comparison of models with different imaging modes as input. U: grayscale mode. D: doppler mode. E: elastography.

| Model | Acc (slice) | Acc (lesion) | AUC (slice) | AUC (lesion) |
|---|---|---|---|---|
| Res3D_U | 69.32 % | 75.76 % | 0.81 | 0.76 |
| Res3D_UD | **76.10 %** | 79.17 % | **0.83** | **0.86** |
| Res3D_UDE | 71.20 % | **82.00 %** | 0.81 | 0.83 |

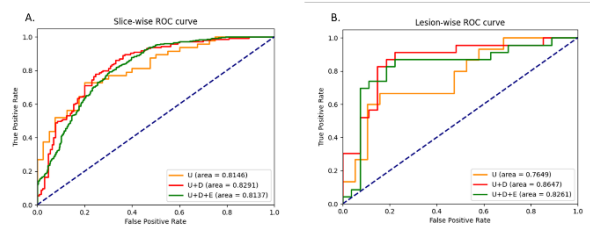

Figure 4. Comparison of models with different imaging modes as input. (a) Slice-wise ROC curve. (a) Lesion-wise ROC curve. U: grayscale mode. D: doppler mode. E: elastography.

### 5. Conclusion

In the present study, the Res3D_UDE we proposed achieved an accuracy of 82.00% and an AUC of 0.83 for lesion classification on EBUS video. By comparing different input data modes, we can find that the accuracy improved when different imaging modes were added, while physicians also need to combine the characteristics of the lesions in different modes to improve the diagnosis rate. Compared with previous studies, our system has several advantages. First, in clinical situation, a certain degree of ultrasonic noise is inevitable. Our training and validation data does not filter the noise artificially, so the model has a certain degree of noise resistance. And there is no need to manually select data. Second, we used a 3D convolutional neural network to obtain information in time dimension in the video, which effectively improved the overall accuracy. After being evaluated by physician, we believe the system with such accuracy rate can be applied to clinical practice in the future, while no extra time is needed to select data. The system not only provided confidences score in real time for clinical reference but can also provide guidance to inexperienced physicians.